\input{aipcheck}
\documentclass[draft, numberedheadings]{aipproc}
\layoutstyle{6x9}

  \newcommand{\beq}{\begin{equation}}
  \newcommand{\eeq}{\end{equation}}
  \newcommand{\beql}[1]{\begin{equation}\label{eq:#1}}

  \newcommand{\beqa}{\begin{eqnarray}}
  \newcommand{\eeqa}{\end{eqnarray}}
  \newcommand{\beqas}{\begin{eqnarray*}}
  \newcommand{\eeqas}{\end{eqnarray*}}
  \newtheorem{Theorem}{Theorem}

 \newcommand{\qed}{{\em QED}}

  \newcommand{\R}{{\bf R}}

  \newcommand{\da}{\dagger}
  
  \newcommand{\de}{\delta}
  \newcommand{\ep}{\epsilon}

  \newcommand{\mb}{\mbox}

  \newcommand{\ph}{\phi}
 \newcommand{\ps}{\psi}
  \newcommand{\rh}{\rho}
  \newcommand{\si}{\sigma}

  \newcommand{\De}{\Delta}

  \newcommand{\And}{\wedge}
  
 \newcommand{\Tr}{\mbox{\rm Tr}}
  
  \newcommand{\bx}{{\bf x}}
  \newcommand{\by}{{\bf y}}
  
  \newcommand{\eq}[1]{(\ref{eq:#1})}

\newcommand{\bra}[1]{\left\langle#1\right|}
\newcommand{\ket}[1]{\left|#1\right\rangle}
\newcommand{\ketbra}[1]{\ket{#1}\!\!\bra{#1}}

\newcommand{\bracket}[1]{\left\langle#1\right\rangle}

  \newcommand{\bM}{{\bf M}}

  \newcommand{\bP}{{\bf P}}
  
  \newcommand{\bS}{{\bf S}}

  \newcommand{\cC}{{\cal C}}

  \newcommand{\cH}{{\cal H}}

  \newcommand{\cK}{{\cal K}}

\newcommand{\commutes}{\leftrightarrow} 
\newcommand{\ppsi}{\ket{\psi}}
\newcommand{\pps}{\ket{\psi}}
\newcommand{\bbracket}[3]{\left\langle#1\left|#2\right|#3\right\rangle}
\newcommand{\is}{\|\,}
\renewcommand{\rh}{\pps}
\renewcommand{\ps}{\pps}
\renewcommand{\ph}{\ket{\phi}}
\newcommand{\M}{\bM}
\renewcommand{\si}{\ket{\xi}}

\begin{document}

\title{Universal Uncertainty Principle, 
Simultaneous Measurability, and Weak Values}


\author{Masanao Ozawa}{
  address={Graduate School of Information Science,
Nagoya University,
Chikusa-ku, Nagoya, 464-8601, Japan}
}

\begin{abstract}
In the conventional formulation,
it is broadly accepted that simultaneous measurability and 
commutativity of observables are equivalent. 
However,  several objections have been claimed that there are cases in which 
even nowhere commuting observables can be measured simultaneously.
Here, we outline a new theory of simultaneous measurements based on a
state-dependent formulation, in which nowhere commuting observables
are shown to have simultaneous measurements in some states, so that 
the known objections to the conventional theory are theoretically justified.
We also discuss new results on the relation between weak values and output  
probability distributions of simultaneous measurements.
\end{abstract}

\maketitle


\section{Introduction}

In the conventional formulation of quantum mechanics, it is broadly accepted 
that simultaneous measurability and commutativity of observables are equivalent. 
However, there have been known several objections on this statement.
To eliminate an intermediate case in which the two observables commute on a
subspace, we say that two observables are nowhere commuting, if they have
no common eigenstate.
Then, the objections generally state that there are cases in which even nowhere 
commuting observables are simultaneously measurable. 
In one case mentioned by Heisenberg \cite{Hei30},
if at time 0 the object is prepared in
an eigenstate of $A$ and the observer actually measures the value of another
observable $B$ nowhere commuting with $A$, then the observer can know 
both the values of two observables $A$ and $B$ at time 0.
In the other case mentioned by
Schr\"{o}dinger \cite{Sch35} following Einstein-Podolsky-Rosen (EPR) \cite{EPR35},
if at time  0 two particles, I and II,  are in the EPR state, 
then the observer can measure the position of particle I and the momentum of
particle II simultaneously at time 0, and the observer knows the momentum of
particle I at time 0 due to the EPR correlation as well as the position of
particle I at time 0 by the direct measurement.
The above objections have been known for long time.
However, no rigorous and general  theory of simultaneous measurements 
has been known that puts those cases in the right place.

In this paper we shall outline a general theory of simultaneous measurements
based on rigorous theory of quantum measurements
\cite{84QC}--\cite{08QLM}.
In this general theory, simultaneous measurements of nowhere commuting observables
are shown to be possible and the above two cases are included as typical cases.
We also discuss several topics such as 
the relation between simultaneous measurability and commutativity in a
state-dependent formulation,
the characterization of simultaneously measurable observables, and 
the universal uncertainty principle for approximate simultaneous measurements,
including new results on 
the relation between weak values and output  probability distributions
of simultaneous measurements.
Proofs of new theorems will be published elsewhere.

\section{State-Dependent Commutativity}

In this paper, we consider a quantum system $\bS$ described by a finite
dimensional Hilbert space $\cH$.  A self-adjoint operator on $\cH$ is called
an {\em observable} and a unit vector in $\cH$ is called a {\em state}.
For any observable $A$, we denote by $E^{A}(a)$ the projection of $\cH$ 
onto the subspace $\{\ppsi\in\cH\mid A\ppsi=a\ppsi\}$, 
which is the eigensubspace of $A$ belonging to $a$, if $a$ is an eigenvalue, 
or the 0-dimensional subspace $\{0\}$, otherwise.
The {\em infimum} of two projections $E$ and $F$ is the projection $E\And F$ 
onto the intersection of ranges of $E$ and $F$.  Thus, the projection $E^{A}(a)\And E^{B}(b)$
is the projection onto the subspace  $\{\ppsi\in\cH\mid \mb{$A\ppsi=a\ppsi$ and $B\ppsi=b\ppsi$}\}$.

In the conventional formulation,
it has been well accepted that two observables are simultaneously measurable
if and only if they commute.
However, this statement is based on the state-independent formulation.
Several objections have been claimed that there are cases in which non-commuting
observables can be measured simultaneously in particular states.
Here, we consider the state-dependent formulation of simultaneous measurability,
in which we shall determine the condition for two observables to be simultaneously 
measurable in a given state.
In order to establish this, we start with introducing the state-dependent
notion of commutativity of observables.

An {\em $n$-dimensional probability distribution in $x_1,\ldots,x_n$} is
a family $\{p(x_1,\dots,x_n)\mid (x_1,\dots,x_n)\in\R^{n}\}$ 
of non-negative numbers such that
$\sum_{x_1,\ldots,x_n} p(x_1,\dots,x_n)=1$,
where $\R$ denotes the set of real numbers.
A {\em joint probability distribution (JPD)}  of two observables $A, B$ 
in a state $\ppsi$ is a 2-dimensional
probability distribution (PD) $\Pr\{A=a,B=b\is\pps\}$ in $a,b$ satisfying the relation
\beqas
\bbracket{\psi}{f(A,B)}{\psi}=\sum_{a,b} f(a,b)\,\Pr\{A=a,B=b\is\pps\}
\eeqas
for any polynomial $f(A,B)$ in $A$ and $B$.
The JPD is inherently a state-dependent notion and plays a central role
in our theory. 
Gudder \cite{Gud68} obtained the following characterization of the existence of JPDs. 

\begin{Theorem}[Gudder 1968]\label{th:Gudder-1}
For any observables $A$ and $B$ and any state $\ppsi$,
the following conditions are equivalent.

(i) There exists a JPD of $A,B$ in $\ppsi$.

(ii) For any $a, b\in\R$ we have $[E^{A}(a),E^{B}(b)]\ppsi=0$.

(iii) The state $\pps$ is a superposition of common eigenstates of $A$ and $B$.

(iv) The family $p(a,b)=\bbracket{\psi}{E^{A}(a)\And E^{B}(b)}{\psi}$
is a 2-dimensional probability distribution in $a,b$, or equivalently
$\sum_{a,b}\bbracket{\psi}{E^{A}(a)\And E^{B}(b)}{\psi}=1$.
\end{Theorem}

We say that two observables $A$ and $B$ {\em commute} in a state $\ppsi$,
in symbols $A\commutes_{\ppsi} B$, if one of the above conditions holds.
We also say that {\em two observables $A$ and $B$ commute nowhere},
if they commute in no states, or equivalently if they have no common eigenstates.
Spin components in different directions commute nowhere. 

\section{Joint Quasiprobability}

Since JPDs do not always exist, it is convenient to introduce mathematical
notions that exist for any pair of observables and reduce to the JPD 
for commuting pairs.
Here, we discuss two versions, weak and strong, of JPDs.

The {\em strong joint quasiprobability distribution} of observables $A$ and $B$  
in a state $\ppsi$ is defined by 
 $$
 {\rm Pr}_{S}\{A=a,B=b\is\pps\}=\bbracket{\psi}{E^{A}(a)\And E^{B}(b)}{\psi}
 $$
 for any $a,b\in\R$.
 The {\em weak joint quasiprobability distribution} of observables $A$ and $B$
 in a state $\ppsi$ is defined by
 $$
 {\rm Pr}_{W}\{A=a,B=b\is\pps\}=\bbracket{\psi}{E^{B}(a)E^{A}(b)}{\psi}
 $$
 for any $a,b\in\R$; if the above relation defines a 2-dimensional
 probability distribution, we call it the {\em weak joint probability distribution}.
 The notion of strong joint quasiprobability distribution was originated by
 Birkhoff and von Neumann \cite{BvN36}, who assigned 
the projection $E^{A}(a)\And E^{B}(b)$ to  the proposition "$A=a$ and $B=b$.'' 
 The weak joint quasiprobability distribution was introduced by Kirkwood \cite{Kir33}.
 The above notions reduce to the JPD in the case where $A$ and $B$ commute in
 $\ps$ as follows.
 \begin{Theorem}
 \label{th:JPD}
  For any observables $A$ and $B$ and any state $\ppsi$, observables 
 $A$ and $B$ commute in $\ps$ if and only if the strong and the weak
 joint quasiprobability distributions coincide, i.e., 
 $$
{\rm Pr}_{S}\{A=a,B=b\is\ppsi\}
={\rm Pr}_{W}\{A=a,B=b\is\ppsi\}
$$ 
 for any $a,b\in R$.
\end{Theorem}

From Theorem \ref{th:Gudder-1} (iv), the strong joint quasiprobability distribution of
$A,B$ in $\ps$ is a 2-dimensional probability distribution 
if and only if $A$ and $B$ commute in $\ps$.
However, the weak joint quasiprobability distribution of
$A,B$ in $\ps$ can be a 2-dimensional probability distribution 
even if $A$ and $B$ do not commute in $\ps$ as discussed in Section \ref{se:SM}.

\section{Conditional Quasiprobability}

In probability theory, the conditional probability is defined as the ratio of 
the joint probability relative to the marginal probability and 
the conditional expectation is defined as the expectation with respect 
to the conditional probability distribution. 
Analogously, conditional probability distribution and conditional expectation
of quantum observables are given as follows.
If observables $A$ and $B$ commute in a state $\ps$
and satisfy $\Pr\{B=b\is\ppsi\}>0$, we naturally introduce 
the {\em conditional probability distribution} of $A$ given 
$B=b$ as 
$$
\Pr\{A=a|B=b\is\ppsi\}=\frac{\Pr\{A=a,B=b\is\ppsi\}}{\Pr\{B=b\is\ppsi\}}.
$$
Then, the {\em conditional expectation} of $A$ given $B=b$ is defined
by 
$$
{\rm Ex}\{A|B=b\is\rh\}=\sum_{a} a {\rm Pr}\{A=a|B=b\is\rh\}.
$$

Now, we assume $\Pr\{B=b\is\ppsi\}>0$.
The strong and weak versions of conditional quasiprobability is introduced without
assuming $A\commutes_{{\pps}}B$ as follows.  
The {\em strong conditional quasiprobability distribution} of $A$ given $B=b$ is 
defined by
$$
{\rm Pr}_{S}\{A=a|B=b\is\rh\}=\frac{{\rm Pr}_{S}\{A=a,B=b\is\ps\}}{\Pr\{B=b\is\ps\}}.
$$
Since the weak joint quasiprobability distribution is not symmetric in $A$ and $B$, 
there are two different notions of conditional quasiprobability corresponding to
preselection and postselection.  Since they can be easily convert each other,
in the following we introduce only the postselection version.
The {\em weak postconditional quasiprobability distribution} of $A$ given $B=b$ is defined by
$$
{\rm Pr}_{W}\{A=a|B=b\is\rh\}=\frac{{\rm Pr}_{W}\{A=a,B=b\is\ps\}}{\Pr\{B=b\is\ps\}}.
$$
Now, weak and strong versions of conditional quasiexpectations are in order.
The {\em strong conditional quasiexpectation} 
and the {\em weak postconditional quasiexpectation} of $A$ given $B=b$ are
defined, respectively, as
\beqa
{\rm Ex}_{S}\{A|B=b\is\rh\}&=&\sum_{a} a {\rm Pr}_{S}\{A=a|B=b\is\rh\},\\
{\rm Ex}_{W}\{A|B=b\is\rh\}&=&\sum_{a} a {\rm Pr}_{W}\{A=a|B=b\is\rh\}.
\eeqa
It is easy to see that 
$$
{\rm Ex}_{W}\{A|B=b\is\rh\}=
\frac{\bracket{\psi|E^{B}(b)A|\psi}}{\bracket{\psi|E^{B}(b)|\psi}},
$$
and 
from Theorem \ref{th:JPD},
we have
$$
{\rm Ex}\{A|B=b\is\rh\}={\rm Ex}_{S}\{A|B=b\is\rh\}={\rm Ex}_{W}\{A|B=b\is\rh\}
$$
if  $A\commutes_{{\pps}}B$.

\section{Weak Values} 

The notion of weak values introduced 
by Aharonov-Albert-Vaidman \cite{AAV88} can be mathematically formulated
as follows. 
For an observable $A$ and two states $\ket{\psi_i}$ and $\ket{\psi_f}$, 
the weak value of $A$ with preselected state  $\ket{\psi_i}$ and postselected
state $\ket{\psi_f}$ is defined by
$$
A_{w}=\frac{\bbracket{\psi_f}{A}{\psi_i}}{\bracket{\psi_f|\psi_i}}.
$$

As pointed out by Steinberg \cite{Ste95,Ste95b}, the formal definition of 
the weak value coincides with the probability theoretical notion of 
weak postconditional quasiexpectation, which as shown above is naturally defined 
based on the Kirkwood joint quasiprobability \cite{Kir33}.
\begin{Theorem}[Steinberg 1995]
For observable $B=\sum_{b}b\ketbra{B=b}$, 
the weak value $A_{w}$ of an observable $A$ with preselected state  $\ket{\psi_i}=\ket{\psi}$ and postselected
state $\ket{\psi_f}=\ket{B=b}$ coincides with 
the weak postconditional quasiprobability distribution of $A$ given $B=b$,
i.e., 
$$
A_{w}={\rm Ex}_{W}\{A|B=b\is\ppsi\}.
$$
\end{Theorem}
 
\section{State-Dependent Identity}
In quantum mechanics, one of the fundamental questions is to answer 
when two observables are considered to have the same value.
However, this question has eluded a satisfactory solution for long time.
In the recent work \cite{05PCN,06QPC}, the present author found 
a plausible solution to this problem based on the weak joint quasiprobability
distribution.

We say that two observables $A$ and $B$ are {\em identically correlated} (or 
{\em perfectly correlated}) in a state $\rh$,
in symbols $A=_{\rh}B$,
if 
$$
 {\rm Pr}_{W}\{A=a,B=b\is\pps\}=0,
$$
provided $a\not=b$.
For observables $A,B$ and a state $\ps$  
the cyclic subspace $\cC(A,B,\ps)$ generated by $A,B,\ps$ 
is defined  by
 $$
 \cC(A,B,\ps)=\{f(A,B)\ps\mid \mb{$f$ is a polynomial}\}.
 $$
 We shall define  $\cC(A,\ps)$ by
 $\cC(A,\ps)=\cC(A,I,\ps)$.
The following theorem characterizes the identical correlation between two observables.

\begin{Theorem}
\label{th:IDC}
For any observables $A,B$ and state $\ps$,
the following conditions are equivalent.

(i) $A=_{\ps}B$.

(ii) $A\commutes_{\ps}B$ and $\sum_{a}{\rm Pr}\{A=a,B=a\|\ps\}=1$.

(iii) $\ps$ is a superposition of common eigenstates of 
$A$ and $B$ with common eigenvalues.

(iv) $A=B$ on $\cC(A,\ps)$.

(v) $E^{A}(x)=E^{B}(x)$  on $\cC(A,\ps)$ for all $x$.

(vi) $\Pr\{A=x\is\ph\}=\Pr\{B=x\is\ph\}$ for all $x$ and 
$\ph\in\cC(A,\ps)$.

(vii) $\|A\ph-B\ph\|=0$ for all $\ph\in\cC(A,\ps)$.

(viii)  $\sum_{a}{\rm Pr}_{S}\{A=a,B=a\|\ps\}=1$.

(ix) $E^{A}(a)_{w}=\de_{a,b}$ if
$\Pr\{B=b\is\ps\}>0$, where $\ket{\psi_i}=\ppsi$ and $\ket{\psi_f}=\ket{B=b}$.

(x) ${\rm Pr}_{S}\{A=a,B=b\is\ps\}=
{\rm Pr}_{W}\{A=a,B=b\is\ps\}=\de_{a,b}\Pr\{A=a\is\ps\}$.
\end{Theorem}

An important property of quantum identical correlation is transitivity
\cite{05PCN,06QPC}, whereas commutativity does not have this property.

\begin{Theorem} The relation
$=_{\ppsi}$ is an equivalence relation among observables.
In particular, if $A=_{\ppsi}B$ and $B=_{\ppsi}C$,
then $A=_{\ppsi}C$.
\end{Theorem}

\section{Measuring Processes}
A {\em measuring process} for $\cH$ is defined to be a quadruple
$(\cK,\si,U,M)$ consisting of a Hilbert space $\cK$, 
a state $\si$ in $\cK$, 
a unitary operator $U$ on $\cH\otimes\cK$,
and an observable $M$ on $\cK$ \cite{84QC}.
The measuring process $\M=(\cK,\si,U,M)$ describes a
measurement carried out by an interaction, called the {\em measuring interaction}, 
between the system $\bS$ described by $\cH$ 
and the {\em probe} system $\bP$ described by $\cK$ 
that is prepared in the state $\si$ just before the measuring interaction.
The unitary operator $U$ describes the time evolution during the measuring 
interaction, say, from time 0 to $\De t$.
For any observables $A,B$ on $\cH$ and $M$ on $\cK$,  we write
$A(0)=A\otimes I$, 
$B(0)=B\otimes I$, 
$M(0)=I\otimes M$,
$A(\De t)=U^{\da}(A\otimes I)U$,
$B(\De t)=U^{\da}(B\otimes I)U$, and
$M(\De t)=U^{\da}(I\otimes M)U$.
An $n$-dimensional POVM is a family of positive operators 
$\Pi(x_1,\ldots,x_n)$ on $\cH$ with
$(x_1,\ldots,x_n)\in\R^{n}$ such that $\sum_{x_1,\ldots,x_n}\Pi(x_1,\ldots,x_n)=I$.
For measuring process $\M$, the relation
\beqa\label{eq:POVM}
\Pi(x)
=\Tr_{\cK}[E^{M(\De t)}(x)(I\otimes\ketbra{\xi})].
\eeqa 
defines a 1-dimensional POVM called the POVM of  $\M$.
The outcome of measuring process $\M$ is obtained by
measuring the observable $M$, called the {\em meter observable},  
in the probe at the time just
after the measuring interaction.  Thus, the {\em output probability distribution}
$\Pr\{\bx=x\is\ps\}$,
the probability distribution 
of the output variable $\bx$ of this measurement on an arbitrary input state $\ps$, 
is given by the POVM of measuring process $\M$ as
\beqa
\Pr\{\bx=x\is\rh\}=\Pr\{M(\De t)=x\is\pps\ket{\xi}\}=\bracket{\psi|\Pi(x)|\psi}.
\eeqa

\section{Universally valid uncertainty principle}
A {\em simultaneous measuring process} is a 6-tuple $\M=(\cK,\si,U,M,f,g)$ consisting of 
a measuring process $(\cK,\si,U,M)$ and a pair of real functions $f,g$.  
For observables $A, B$ and a state $\ps$, the {\em root-mean-square (rms) errors}
$\ep(A)$ and $\ep(B)$ of $\M$
for the $A$-measurement and the $B$-measurement, respectively, in $\ps$ is
defined by
\beqa
\ep(A)&=&\|N(A)\pps\ket{\xi}\|,\\
\ep(B)&=&\|N(B)\pps\ket{\xi}\|,
\eeqa
where the {\em noise operators} $N(A),N(B)$ are given by
$N(A)=f(M(\De t))-A(0)$ and $N(B)=g(M(\De t))-B(0)$.
Then, those errors always satisfy the following universally valid relation \cite{03UVR,03UPQ}
\beqa\label{eq:UUP}
\ep(A)\ep(B)+\ep(A)\sigma(B)+\sigma(A)\ep(B)\ge\frac{1}{2}\left|\bracket{\psi|[A,B]\psi}\right|,
\eeqa
where $\sigma(A), \sigma(B)$ are the standard deviations of $A,B$ in $\ps$.
If the {\em mean errors} $\bracket{\xi|\bracket{\psi|N(A)|\psi}|\xi}$ and 
$\bracket{\xi|\bracket{\psi|N(B)|\psi}|\xi}$
are independent of the initial state $\si$ of the probe, we have \cite{03UVR,03UPQ}
\beqa\label{eq:HUP}
\ep(A)\ep(B)\ge\frac{1}{2}\left|\bracket{\psi|[A,B]|\psi}\right|,
\eeqa
a form suggested originally by Heisenberg's gamma-ray microscope thought 
experiment \cite{Hei27}.
The last relation is obviously not universally valid, since $\ep(A)=0$ implies
$\ep(B)\sim\infty$ in a state with $\bracket{\psi|[A,B]|\psi}\not=0$, whereas the
new relation, \eq{UUP}, concludes the new constraint $\sigma(A)\ep(B)\ge
\frac{1}{2}\left|\bracket{\psi|[A,B]|\psi}\right|$ \cite{03UVR,03UPQ}.
Hall \cite{Hal04} pointed out the role of prior information in the violation of 
\eq{HUP}.

\section{Simultaneous measurability}
\label{se:SM}

For observables $A, B$ and a state $\ps$, a {\em simultaneous measurement} of $A,B$
in $\ps$ is a simultaneous measuring process $(\cK,\si,U,M,f,g)$ satisfying
\beqas
f(M(\De t))&=_{\pps\ket{\xi}}&A(0),\\
g(M(\De t))&=_{\pps\ket{\xi}}&B(0).
\eeqas
The above condition is equivalent to the condition that 
the measuring process $(\cK,\si,U,f(M))$ measures $A$ and that 
the measuring process $(\cK,\si,U,g(M))$ measures $B$. 
Two observables $A,B$ are said to be {\em simultaneously measurable} in a state $\ppsi$
if there exists a simultaneous measurement of $A,B$ in $\ppsi$.

From Theorem \ref{th:IDC} (vii), any simultaneous measurement of $A,B$ satisfies
$\ep(A)=\ep(B)=0$, so that $\bracket{\psi|[A,B]|\psi}=0$.
Moreover, any simultaneous measurement of $A,B$ can be easily modified to be
a simultaneous measurement of $E^{A}(a),E^{B}(b)$, so that we also have
$\bracket{\psi|[E^A(a),E^{B}(b)]|\psi}=0$ for all $a,b$; however, this does not
imply that $A$ and $B$ commute in $\ps$, since the latter condition is equivalent
to $\bracket{\psi|[E^A(a),E^{B}(b)]^2|\psi}=0$.
Thus, a simultaneous measurement of nowhere commuting observables $A$ and $B$
is possible only in the case where $\bracket{\psi|[E^A(a),E^{B}(b)]|\psi}= 0$ for all 
$a,b$ but $\bracket{\psi|[E^A(a),E^{B}(b)]^2|\psi}\ne 0$ for some $a,b$.

\sloppy
The {\em joint output probability distribution} of 
a simultaneous measuring process $(\cK,\si,U,M,f,g)$ is the joint probability
distribution of the output variables $\bx,\by$ defined by
\beqa
\Pr\{\bx=x, \by=y\is\ps\}=\Pr\{f(M(\De t))=x,g(M(\De t))=y\is\ps\si\}.
\eeqa
Then, we have
$$
\Pr\{\bx=x, \by=y\is\ps\}=\sum_{u: x=f(u),y=g(u)}\bracket{\psi|\Pi(u)|\psi}.
$$

Simultaneous measurability and  commutativity are not
equivalent notion under the state-dependent formulation, as
the following theorem clarifies \cite{11QRM}.

\begin{Theorem}\label{th:main}
(i) Two observables $A,B$ commute in a state $\rh$ if and only if
there is a 2-dimensional  POVM $\Pi$ such that 
\beqa
\sum_y\Pi(x,y)&=&E^{A}(x)\quad\mbox{on}\quad\cC(A,B,\rh),
\label{eq:i-1}\\
\sum_x\Pi(x,y)&=&E^{B}(y)\quad\mbox{on}\quad\cC(A,B,\rh).
\label{eq:i-2}
\eeqa

(ii) Two observables $A,B$ are
simultaneously measurable in a state $\rh$
 if and only
if there is a 2-dimensional  POVM $\Pi$ such that 
\beqa
\sum_y\Pi(x,y)&=&E^{A}(x)\quad\mbox{on}\quad\cC(A,\rh),
\label{eq:ii-1}\\
\sum_x\Pi(x,y)&=&E^{B}(y)\quad\mbox{on}\quad\cC(B,\rh).
\label{eq:ii-2}
\eeqa

(iii) Two observables  are
simultaneously measurable in a state $\rh$
if they commute in $\rh$. 
\end{Theorem}

In the conventional theory, the joint output probability distribution
of the simultaneous measurement of two commuting 
observables $A$ and $B$ in a state $\psi$ 
is shown to be given by their joint probability distribution.  
However, if $A$ and $B$ are not commuting, this relation is no longer
meaningful.  Instead, we can show that the joint output probability distribution
of the simultaneous measurement is always given by the weak joint
probability distribution.

\begin{Theorem}
For any simultaneous measurement $(\cK,\si,U,M,f,g)$ of observables $A,B$
in a state $\ps$, the joint output probability distribution does not depend on 
the measuring process but always coincides with the weak joint quasiprobability 
distribution, i.e., 
$$
\Pr\{\bx=x, \by=y\is\pps\}={\rm Pr}_{W}\{A=x,B=y\is\pps\}.
$$
\end{Theorem}

\section{Characterizations}

In this section, we collect results on 
characterizations of simultaneous measurements
of nowhere commuting observables.

The conventional relation between simultaneous measurability and
commutativity is recovered as follows.

\begin{Theorem}
Observables $A$ and $B$ are simultaneously measurable in every state $\rh$
if and only if $A$ and $B$ commute on $\cH$.
\end{Theorem}

The following theorems show that we can simultaneously measure 
two nowhere commuting observables  \cite{11QRM}.
   
\begin{Theorem}\label{th:13}
In any Hilbert space,
every pair of observables are simultaneously measurable
in any eigenstate of either observable.
\end{Theorem}

\begin{Theorem}
In any Hilbert space with dimension more than 3,
there are nowhere commuting observables
that are simultaneously measurable in a state that is
not an eigenstate of either observable.
\end{Theorem}

In the case where $\dim(\cH)=2$, the following characterization holds.

\begin{Theorem}
Assume $\dim(\cH)=2$.
The following conditions are equivalent.

(i) $A$ and $B$ are simultaneously measurable in $\rh$.

(ii) ${\rm Pr}_{W}\{A=a,B=b\|\rh\}\ge 0$ for all $a,b$.

(iii) Either $A$ and $B$ commute in $\rh$, or
$\rh$ is an eigenstate of $A$ or $B$.
\end{Theorem}

It is an open problem whether the equivalence between conditions 
(i) and (ii) above holds in arbitrary dimensions.


\begin{theacknowledgments}
This work was supported in part by KAKENHI 21244007.
\end{theacknowledgments}

\bigskip

{\small (Work presented at the 10th International Conference on Quantum Communication, Measurement and Computation (QCMC 2010), Brisbane, Australia, 19 July 2010 as 
one of award talks.)}

\end{document}